\documentclass[11pt, fleqn]{article}
\usepackage{amsmath, amssymb,amsthm,cite}
\usepackage[dvips]{graphics,color}
\usepackage {graphicx}

\title{Symmetries, conservation laws and exact solutions of static plasma equilibrium systems in three dimensions}
\author{ 
Alexei F. Cheviakov\footnotemark[1], \\
{\small \emph{Department of Mathematics, University of British Columbia, Vancouver, V6T 1Z2 Canada}}\\
 Stephen C. Anco\footnotemark[2]\\
\small \emph{Department of Mathematics, Brock University, St. Catharines, Canada}\\}

\newtheorem{theorem}{Theorem}

\newtheorem{corollary}{Corollary}
{\theoremstyle{definition}
\newtheorem{remark}{Remark}

\newcommand{\vspacebefore}{\raisebox{0ex}[2.5ex][0ex]{\null}}

\def\const{\hbox{\rm const}}

\def\grad{\mathop{\hbox{\rm grad}}}

\def\div{\mathop{\hbox{\rm div}}}
\def\curl{\mathop{\hbox{\rm curl}}}
\def\vec#1{{\boldsymbol{\rm #1}}}  
\def\tens#1{{\mathbb {#1}}}
\def\abs#1{|\vec{#1}|}
\def\hook{\lrcorner}
\def\flux{{\mathcal F}}

\begin{document}
\footnotetext[1]{Electronic mail: alexch@math.ubc.ca} \footnotetext[2]{Electronic mail: sanco@brocku.ca}

\maketitle
\begin{abstract}
For static reductions of isotropic and anisotropic Magnetohydrodynamics plasma equilibrium models, a complete classification
of admitted point symmetries and conservation laws up to first order is presented. It is shown that the symmetry algebra for the isotropic
equations is finite-dimensional, whereas anisotropic equations admit infinite symmetries depending on a free function
defined on the set of magnetic surfaces. A direct transformation is established between isotropic and anisotropic equations,
which provides an efficient way of constructing new exact anisotropic solutions. In particular, axially and helically
symmetric anisotropic plasma equilibria arise from classical Grad-Shafranov and JFKO equations.
\end{abstract}

\bigskip
\textbf{PACS Codes:} 05.45.-a , 02.30.Jr, 02.90.+p, 52.30.Cv.

\bigskip
\textbf{Keywords:} Plasma equilibrium; Symmetries; Conservation laws; Exact solutions; Grad-Shafranov equation.


\renewcommand{\baselinestretch}{1}\small\normalsize
\renewcommand{\theequation}{\arabic{section}.\arabic{equation}}

\section{Introduction.}
\smallskip
Systems of isotropic Magnetohydrodynamics (MHD) and anisotropic Chew-Goldberger-Low (CGL) plasma equations, in particular,
their equilibrium reductions, are used for description of plasmas in controlled thermonuclear fusion research, geophysics
and astrophysics (Earth magnetosphere, star formation, solar activity), and laboratory and industrial applications
\cite{fusion1,CGL,bisk}.

MHD and CGL systems, as well as their equilibrium versions, are essentially nonlinear systems of partial differential
equations in 3D space. Knowledge of physically meaningful exact solutions and analytical properties of these systems (such
as symmetries, conservation laws, stability criteria, etc.) is important for understanding the core properties of the
underlying physical phenomena, for modelling, and for the development of appropriate numerical methods.

Common ways of finding exact analytical solutions to such systems include reduction by a symmetry group (similarity
solutions), the use of symmetry transformations to generate new solutions from known ones, and the use of mappings from
solutions of other equations. The first approach applied to axially and helically symmetric static MHD configurations has
yielded the well-known Grad-Shafranov \cite{BraggH,GradRubin,Shafr} and JFKO \cite{jfko} equations, and led to several
classes of exact solutions (e.g. \cite{ob_prl,ball2,ball4,jets4}). A different approach that makes use of equilibrium
solution topology (general existence of 2D magnetic surfaces) was used in {\cite{afc_prl}}.

A symmetry of a system of PDEs is any transformation of its solution manifold into itself. A symmetry thus maps any
solution to another solution of the same system. Several types of symmetries, such as continuous (point, contact,
higher-order, nonlocal) Lie groups of symmetries and discrete symmetries, can be obtained algorithmically (e.g.
\cite{olv,BlumanAncoBook,BC1,BC2,hydon2}). In particular, using Lie's algorithm for solving symmetry determining equations,
one can discover one-parameter, multi-parameter, and infinite-dimensional symmetry groups. Symmetries are used as
transformations that yield new solutions and new conservation laws of differential equations from known ones, and also for
finding particular symmetry-invariant solutions. Knowledge of symmetries is essential to answer the question about the
possibility of mapping a given PDE system into a target PDE system or a class of PDE systems.

An important complement to the full symmetry structure of a PDE system is knowledge of its conservation law structure. Local
conservation laws contain important information about physical properties of a model under consideration, and provide
conserved norms used in analysis of solutions and also in development of numerical methods. Conservation laws can be found
algorithmically by a direct construction method in terms of multipliers that satisfy determining equations related to the
adjoint of the symmetry ones \cite{AB97,AB02p12}, without the need for any Lagrangian. In particular, this method allows one
to by-pass all limitations of Noether's theorem.

The paper is organized as follows. In Section \ref{sec:systems}, we describe the static plasma models under consideration,
as well as their basic properties, and state the transformation that relates the two models. In Section
\ref{sec:Sym_analys}, we classify and compare point symmetries of MHD and CGL static plasma equilibrium systems. In
particular, we demonstrate that the static CGL system admits an infinite-dimensional symmetry group (which appears to be
related to the infinite-dimensional symmetry group of dynamic CGL equilibrium system \cite{afc_ob}). We derive infinitesimal
and global representations of the admitted symmetry groups, and discuss physical properties and group structure of the
infinite symmetries that arise for the static CGL system. In Section \ref{sec:constr}, we discuss applications of the
infinite symmetry group to construction of exact static anisotropic plasma equilibria. In particular, we show that axial and
helical static anisotropic (CGL) equilibria arise from solutions to conventional Grad-Shafranov and JFKO equations. An
explicit example of an exact solution describing an anisotropic plasma vortex is presented. Finally, in Section
\ref{sec:CL_analys}, we complete the analysis by classifying all conservation laws admitted by static MHD and CGL systems
with multipliers linear in first-order partial derivatives, and discuss their physical meaning. From the comparison of
symmetry and conservation law classifications of these two static plasma equilibrium systems, we establish a direct
transformation (see Theorem \ref{th:CGL_MHD}) between the plasma models (including the equivalence of solution sets.) In Section \ref{sec:further}, we summarize the results presented in this paper in a larger
context of dynamical MHD models.

Symbolic software packages \verb"GeM" for \verb"Maple" \cite{GeM} and \verb"Crack/LiePDE/ConLaw" for \verb"REDUCE"
\cite{wolf_soft} were used for all symmetry and conservation law computations.

\section{Static plasma equilibrium models}\label{sec:systems}
\setcounter{equation}{0}

Equilibrium (time-independent) plasma models with and without flow are used in many physical applications. In particular,
for the purpose of analysis, static plasma equilibrium systems are often considered. On one hand, static equilibrium
equations are much simpler than dynamic ones, and yield to analytical techniques more easily; on the other hand, they are
still nonlinear 3D models, which inherit many properties from full plasma equilibrium models. Static plasma equilibria, even
in a simplified force-free (constant-pressure) setting, are used in astrophysical modelling.

The static MHD equilibrium equations are obtained directly as a static time-independent reduction of the full system of MHD
equations (e.g. {\cite{fusion1}}). They have the form
\begin{equation}\label{eq:PEE}
\curl~{\bf{B}}\times{\bf{B}} = {\rm{grad}}~P , ~~{\div}~{\bf{B}}= 0.
\end{equation}
Here $\bf{B}$ is the vector of the magnetic field induction, and $P$ is plasma pressure. The electric current density is
given by ${\bf{J}}= \curl {\bf{B}}.$

In all static isotropic plasma equilibria (\ref{eq:PEE}), the magnetic field ${\bf{B}}$ is tangent to 2-dimensional magnetic
surfaces $\Psi(x,y,z)=\const$ that span the plasma domain: ${\bf{B}} \cdot \grad \Psi=0.$ The plasma pressure $P=P(\Psi)$ is
constant on magnetic surfaces. In a compact domain, magnetic surfaces are generally tori {\cite{kk}}. When magnetic field
lines are closed loops or go to infinity, magnetic surfaces may not be uniquely defined, and one may specify
$\Psi(x,y,z)=\const$ on each magnetic field line. The only possible case when magnetic surfaces do not exist is
Beltrami-type configurations $\curl{\bf{B}}=\alpha{\bf{B}}$, $\alpha=\const$.

We note the well-known equivalence between the static plasma equilibrium system \eqref{eq:PEE} and the time-independent
Euler's equations of inviscid fluid motion
\begin{equation}\label{eq:Euler}
(\vec{v}\cdot\grad)\,\vec{v} = -\grad\,\pi,~~~\div\,\vec{v}=0,
\end{equation}
with velocity $\vec{v}=\vec{B}$, pressure $\pi=P_0-P-\frac{|\vec{B}|^2}{2}$, and constant density $\rho=1$.

The static time-independent equilibrium of \emph{anisotropic} plasmas is a similar reduction of the CGL equations
\cite{CGL,ThMac, afc_ob}
\begin{equation}\label{eq:APEE}
\left(1-\tau\right) \curl~{\bf{B}}\times{\bf{B}} = {\rm{grad}}~p_\perp +\tau~ {\rm{grad}}\frac{|\vec{B}|^2}{2} +
\bf{B}(\bf{B}\cdot\rm{grad~}\tau), ~~{\div}~{\bf{B}}= 0.
\end{equation}
Here $\tau$ is the anisotropy factor, and pressure $\mathbb{P}$ is a symmetric tensor with two independent parameters
$p_\parallel, p_\perp$:
\begin{equation}
\tau=\frac{p_\parallel-p_\perp}{|\vec{B}|^2},~~~~\tens{P}= p_\perp \tens{I} +\frac{p_\parallel-p_\perp}{|\vec{B}|^2}
\vec{B}\otimes\vec{B},  \label{eq_tau_def}
\end{equation}
where $\tens{I}={\rm diag}(1,1,1)$ is the identity tensor in $\mathbb{R}^3.$

Equations \eqref{eq:APEE} describe equilibria of strongly magnetized or rarified plasmas; $p_\parallel$ denotes pressure
along the strong magnetic field ${\bf B}$, and $p_\perp$ is pressure in the transverse direction. Unlike static isotropic
plasma equilibria, anisotropic plasmas described by \eqref{eq:APEE} in general do not possess 2D magnetic surfaces.

\begin{remark} \label{rem:MHD_IS_CGL}
Static plasma equilibrium systems \eqref{eq:PEE} and \eqref{eq:APEE} arise from Boltzmann and Maxwell equations under
essentially different isotropy assumptions \cite{rose1, krall1, CGL}. In particular, the isotropic model \eqref{eq:PEE} was
derived from Boltzmann equation using expansion in powers of mean free path, while for the anisotropic model the expansion
in powers of ion Larmor radius was used, which implies $p_\parallel \neq p_\perp$. However it is easy to see that for
$\tau=0$ equations \eqref{eq:PEE} and \eqref{eq:APEE} coincide.
\end{remark}

Closure of the anisotropic equilibrium system (\ref{eq:APEE}) requires an equation of state. Note that dynamic equations of
state, such as ``double-adiabatic" equations
\[
\frac{d}{dt}\left(\frac{p_\perp}{\rho|\vec{B}|}\right)=\frac{d}{dt}\left(\frac{p_\parallel|\vec{B}|^2}{\rho^3}\right)=0
\]
suggested in the original CGL paper {\cite{CGL}}, cannot be used since they vanish identically  for all static equilibria.
In this paper we use the equation of state {\cite{afc_ob}}
\begin{equation}
{\bf B} \cdot \grad \tau = 0, \label{eq:tau_cond}
\end{equation}
which implies that the anisotropy factor $\tau$ is constant on magnetic surfaces (more generally, on magnetic field lines).
Below we show that the relation (\ref{eq:tau_cond}) can be rewritten in a form \eqref{eq:tau_cond_new} that is similar to
the equation of state used in numerical modeling of anisotropic magnetosheath plasma in the CGL approximation \cite{erkaev1}
(see also \cite{afc_prl}).

\bigskip
In Sections \ref{sec:Sym_analys} and \ref{sec:CL_analys}, we classify and compare point symmetries and conservation laws
admitted by the isotropic (MHD) static plasma equilibrium model \eqref{eq:PEE} and the anisotropic (CGL) model
\eqref{eq:APEE} with the equation of state \eqref{eq:tau_cond}. The point symmetry classification of the anisotropic model
is found to differ from that of the isotropic model by only one symmetry generator $Y_{\infty}$ \eqref{eq:sym_CGL_inf}
depending on an arbitrary function defined on magnetic surfaces, which describes \emph{infinite symmetries} admitted by the
anisotropic system \eqref{eq:APEE},\eqref{eq:tau_cond}.

Since any solution $({\bf{B}},P)$ of static isotropic (MHD) equilibrium is automatically a solution
$({\bf{B}},{p}_{\perp},{\tau}) = ({\bf{B}},P,0)$ of the static anisotropic (CGL) system \eqref{eq:APEE}, \eqref{eq:tau_cond},
the infinite symmetries $Y_{\infty}$ map each MHD equilibrium solution into a continuum of CGL equilibria. This suggests
that a direct relation may exist between the two static plasma equilibrium systems. Comparison of multipliers and fluxes of
conservation laws admitted by the two plasma equilibrium systems (see Section \ref{sec:CL_analys}) indicates the specific
form of such a transformation: the isotropic equilibrium magnetic field $\vec{B}$ should be proportional to
$\sqrt{1-{\tau}}{\vec{B}}$ of the anisotropic model. This leads to the following important theorem.

\begin{theorem}\label{th:CGL_MHD}
The static anisotropic (CGL) equilibrium system (\ref{eq:APEE}) with equation of state (\ref{eq:tau_cond}) can be written in
the form
\begin{equation}\label{eq:APEE2_1}
\curl\left(\sqrt{1-\tau}{\bf{B}}\right)  \times \left(\sqrt{1-\tau}{\bf{B}}\right) = \grad p,
\end{equation}
\begin{equation}\label{eq:APEE2_2}
\div(\sqrt{1-\tau}{\bf{B}})= 0,
\end{equation}
\begin{equation}\nonumber
{\bf{B}}\cdot\grad\tau=0,
\end{equation}
where $p=p_\perp + \frac{1}{2}\tau {\bf{B}^2}=\frac{1}{2}(p_\parallel +p_\perp)$ is the mean pressure of anisotropic plasma
configuration. This form of the static anisotropic (CGL) equilibrium system (\ref{eq:APEE}), (\ref{eq:tau_cond}) is
equivalent to the static isotropic (MHD) equilibrium system (\ref{eq:PEE}) with pressure $p$, magnetic field
$\sqrt{1-\tau}{\bf{B}}$, for any smooth function $\tau$ constant on magnetic surfaces / magnetic field lines.
\end{theorem}
The proof proceeds by expanding \eqref{eq:APEE2_1}, \eqref{eq:APEE2_2} using vector calculus identities.

\begin{corollary}\label{cor:CGL_MHD}
The 3D domain of every static anisotropic (CGL) equilibrium plasma configuration satisfying \eqref{eq:APEE}, \eqref{eq:tau_cond} is spanned by 2D magnetic surfaces
\begin{equation}
\Psi(x,y,z)=\const \nonumber
\end{equation}
(with the only possible exception being Beltrami-type configurations
$\curl{\sqrt{1-\tau}{\bf{B}}}=\alpha{\sqrt{1-\tau}{\bf{B}}}$, $\alpha=\const$.) In particular, the mean pressure $p=p(\Psi)$
and the anisotropy factor $\tau=\tau(\Psi)$ are constant on magnetic surfaces.

If magnetic surfaces are not uniquely defined (magnetic field lines are closed loops or go to infinity), $\Psi(x,y,z)$ can
have a different constant value on every magnetic field line.
\end{corollary}

This corollary follows from the equivalence of the equations \eqref{eq:APEE2_1} to the static MHD system \eqref{eq:PEE}, for
which this result is well-known.

\bigskip
We also note that since $\tau=\tau(\Psi)$, the equation of state \eqref{eq:tau_cond} can be rewritten as
\begin{equation}
{p_\perp}/{p_\parallel}=1-\left({\tau(\Psi)|\vec{B}|^2}/{p_\parallel}\right). \label{eq:tau_cond_new}
\end{equation}

\section{Point symmetries of plasma equilibrium models}\label{sec:Sym_analys}
\setcounter{equation}{0}

We will classify all point symmetries admitted by static isotropic (MHD) and anisotropic (CGL) systems. In particular, we
seek Lie groups of point symmetries admitted by the system \eqref{eq:PEE}
\begin{equation}\label{eq:p_sym_loc}
\begin{array}{ll}
\vec{x}' = \vec{x} + \epsilon \vec{\xi}(\vec{x},\vec{B},P) + O(\epsilon^2),\\
\vec{B}' = \vec{B} + \epsilon \vec{\eta}(\vec{x},\vec{B},P) + O(\epsilon^2),\\
P' = P + \epsilon \sigma(\vec{x},\vec{B},P) + O(\epsilon^2)\\
\end{array}
\end{equation}
with corresponding infinitesimal generator
\begin{equation} \label{eq:X_MHD}
X  = \vec{\xi}\hook\frac{\partial }{\partial \vec{x}}  + \vec{\eta}\hook\frac{\partial }{\partial \vec{B}}+
\sigma\frac{\partial }{\partial P} ,
\end{equation}
where the hook denotes summation over vector components. Symmetry components $(\vec{\xi}, \vec{\eta}, \sigma)$ satisfy
determining equations given by the invariance of the solution set $(\vec{B}(\vec{x}),P(\vec{x}))$ under $X$:
\begin{equation}\label{eq:MHD_symm_deteq}
\begin{array}{ll}
\curl~{\bf{B}}\times\left(\vec{\eta} -   (\vec{\xi}\cdot \grad) \vec{B}\right) + \curl\left(\vec{\eta} -   (\vec{\xi}\cdot
\grad) \vec{B}\right)\times{\bf{B}}  = {\rm{grad}}~(\sigma -   (\vec{\xi}\cdot\grad) P),
\\
{\div}~\left(\vec{\eta} -   (\vec{\xi}\cdot \grad) \vec{B}\right)= 0
\end{array}
\end{equation}
holding for all static equilibria satisfying \eqref{eq:PEE}.

Similarly,  a symmetry generator of static anisotropic equilibrium system (\ref{eq:APEE}), (\ref{eq:tau_cond}) has the form
\begin{equation} \label{eq:X_MHDD_CGL}
Y  = \vec{\xi}\hook\frac{\partial }{\partial \vec{x}}  + \vec{\eta}\hook\frac{\partial }{\partial \vec{B}}+
\sigma_{\perp}\frac{\partial }{\partial p_{\perp}}+ \nu\frac{\partial }{\partial \tau}
\end{equation}
whose components $(\vec{\xi}, \vec{\eta}, \sigma_{\perp}, \nu)$ are found from corresponding determining equations that
state the invariance of the solution set of the equations (\ref{eq:APEE}), (\ref{eq:tau_cond}) under the action of $Y$
\eqref{eq:X_MHDD_CGL}.

\subsection{Point symmetries of static MHD and CGL plasma equilibrium systems}\label{subs:PointSymmMHD_CGL}

The following theorems describe all Lie point symmetries of static isotropic and anisotropic plasma equilibrium PDE systems.
The proofs are computational and consist of applying Lie's algorithm \cite{olv,BlumanAncoBook} for solving the determining
equations \eqref{eq:MHD_symm_deteq} and their counterpart for the anisotropic case.

\begin{theorem}[Point symmetries of static MHD equations] \label{th:sym_MHD} The point symmetries admitted by the static isotropic plasma equilibrium system
\eqref{eq:PEE} form a nine-dimensional Lie algebra with the following generators:
\begin{itemize}
    \item Killing symmetries (translations and rotations)
         \begin{equation} \label{eq:sym_Killing}
            X_{\rm K} = \vec{\zeta}\hook\frac{\partial}{\partial\vec{x}} + (\vec{B}\cdot\grad)\vec{\zeta}
            \hook\frac{\partial}{\partial\vec{B}};
         \end{equation}
    \item Scalings and dilations
         \begin{equation} \label{eq:sym_ScDil}
            X_{\rm S} = \vec{B}\hook\frac{\partial}{\partial\vec{B}}+2P\frac{\partial }{\partial P},~~~X_{\rm D} = \vec{x}\hook\frac{\partial }{\partial \vec{x}};
         \end{equation}
    \item Pressure shifts
         \begin{equation} \label{eq:sym_Pressure}
            X_{\rm P} = \frac{\partial }{\partial P}.
         \end{equation}
\end{itemize}
Here $\vec{\zeta} =\vec{a}+\vec{b}\times\vec{x}$ is a Euclidean Killing vector; $\vec{a},\vec{b}\in \mathbb{R}^3$ are
constant vectors.

\end{theorem}

\bigskip
\begin{theorem}[Point symmetries of static CGL equations] \label{th:sym_CGL} The static anisotropic plasma equilibrium system \eqref{eq:APEE} with
equation of state \eqref{eq:tau_cond} admits an infinite-dimensional symmetry algebra spanned by the following generators:
\begin{itemize}
    \item Killing symmetries (translations and rotations)
         \begin{equation} \label{eq:sym_KillingCGL}
            Y_{\rm K} =\vec{\zeta}\hook\frac{\partial}{\partial\vec{x}} + (\vec{B}\cdot\grad)\vec{\zeta}
            \hook\frac{\partial}{\partial\vec{B}};
         \end{equation}
    \item Scalings and dilations
         \begin{equation} \label{eq:sym_ScDilCGL}
            Y_{\rm S} = \vec{B}\hook\frac{\partial}{\partial\vec{B}}+2p_{\perp}\frac{\partial }{\partial p_{\perp}},~~~Y_{\rm D} = \vec{x}\hook\frac{\partial }{\partial \vec{x}};
         \end{equation}
    \item Pressure shifts
         \begin{equation} \label{eq:sym_PressureCGL}
            Y_{\rm P} = \frac{\partial }{\partial p_{\perp}};
         \end{equation}
    \item Infinite-dimensional transformations
         \begin{equation} \label{eq:sym_CGL_inf}
            Y_{\infty} = f\vec{B} \hook\frac{\partial}{\partial\vec{B}} - f|\vec{B}|^2\frac{\partial }{\partial p_{\perp}} + 2f(1-\tau) \frac{\partial }{\partial \tau}.
         \end{equation}
\end{itemize}
Here $\vec{\zeta} =\vec{a}+\vec{b}\times\vec{x}$ is a Euclidean Killing vector ($\vec{a},\vec{b}\in \mathbb{R}^3$ are
arbitrary constant vectors), and $f=f\left(p ,~\tau\right)$ is an arbitrary smooth function ($p=p_\perp +\tau \frac{|\bf
B|^2}{2}$ is the mean pressure of anisotropic plasma configuration).
\end{theorem}

The point symmetry classifications of the static MHD and CGL equilibrium systems presented in Theorems \ref{th:sym_MHD} and
\ref{th:sym_CGL} coincide except for the infinite symmetries $Y_{\infty}$ \eqref{eq:sym_CGL_inf} of the CGL system. The
existence of these infinite symmetries underlies the equivalence of the static MHD and CGL equilibrium systems stated in
Theorem \ref{th:CGL_MHD}. Corollary \ref{cor:CGL_MHD} of Theorem \ref{th:CGL_MHD}, in turn, leads to the following
clarification of the form of arbitrary function in the infinite symmetry generator $Y_{\infty}$.

\begin{corollary}\label{cor:CGL_inf_one_func}
The arbitrary function $f=f\left(p,~\tau\right)$ in infinite symmetries \eqref{eq:sym_CGL_inf} of CGL equilibrium equations
can be expressed as a function $f=f(\Psi)$ of a single variable $\Psi=\Psi(\vec{x})$ which enumerates magnetic surfaces (or,
in general, magnetic field lines) of the original plasma configuration.
\end{corollary}

\subsection{Finite form and other properties of symmetry transformations}

Using a standard reconstruction formula (Lie's First Theorem) \cite{BlumanAncoBook,olv}, we find the global Lie groups of
point transformation groups corresponding to symmetry generators \eqref{eq:sym_Killing} - \eqref{eq:sym_CGL_inf}.

\begin{enumerate}
    \item Translations ($X_{\rm K}, Y_{\rm K}$):
    \[
        {\bf x}'={\bf x} + {\bf a}  ~~~~~({\bf a}=(a_1,a_2,a_3)\in \mathbb{R}^3).
    \]

    \item Rotations ($X_{\rm K}, Y_{\rm K}$), parameterized, for example, by Euler angles $\phi,\theta,\psi$:
    \[
    {\bf x}'=A_3 A_2 A_1 {\bf x};~~~{\bf B}'({\bf x}')=A_3 A_2 A_1 {\bf B}({\bf x}),
    \]
    \[
    A_1=\left(\begin{array}{ccc}\cos \phi & \sin \phi & 0\\-\sin \phi & \cos \phi & 0\\0&0&1\end{array}\right);~~
    A_2=\left(\begin{array}{ccc}1&0&0\\0 &\cos \theta & \sin \theta \\0&-\sin \theta & \cos \theta \end{array}\right);~~
    A_3=\left(\begin{array}{ccc}\cos \psi & \sin \psi & 0\\-\sin \psi & \cos \psi & 0\\0&0&1\end{array}\right).
    \]
    \item Scalings ($X_{\rm S}, Y_{\rm S}$):
    \[
        {\bf B}'=a_4 {\bf B},~~ \mathcal{P}'=2 a_4 \mathcal{P}~~~~~(a_4\in \mathbb{R}).
    \]
    ($\mathcal{P} = P$ or $p_\perp$ for MHD and CGL respectively.)
    \item Dilations ($X_{\rm D}, Y_{\rm D}$):
    \[
        {\bf x}'=a_5 {\bf x}~~~~~(a_5\in \mathbb{R}).
    \]
    \item Pressure shifts ($X_{\rm P}, Y_{\rm P}$):
    \[
        \mathcal{P}'=\mathcal{P}+a_6 ~~~~~(a_6\in \mathbb{R}).
    \]
    ($\mathcal{P} = P$ or $p_\perp$ for MHD and CGL respectively.)
    \item Infinite-dimensional transformations ($Y_{\infty}$, anisotropic equilibria only):
    \begin{gather}
        {\bf{B}}'  = M(\Psi)\;{\bf{B}},~~\tau'=1-(1-\tau)M^{-2}(\Psi), \nonumber\\
        p_{\perp}'= p_{\perp} + \frac{{\bf{B}}^2-({\bf{B}}')^2}{2}, ~~ p_{\parallel}'= p_{\perp}' + ({\bf{B}}')^2\left(1-\left(1-\frac{p_{\parallel}-p_{\perp}}{{\bf{B}}^2}\right)M^{-2}(\Psi)\right), \label{eq:finite_form_inf}
    \end{gather}
\end{enumerate}
where $\Psi$ is a function enumerating magnetic surfaces (more generally, magnetic field lines) of the original plasma
configuration $({\bf{B}},p_{\perp},p_{\parallel})$, and the arbitrary function $M(\Psi)$ is related to the function
$f\left(p ,~\tau\right)=f(\Psi)$ in \eqref{eq:sym_CGL_inf}.

It is easy to see why no infinite transformations similar to $Y_{\infty}$ are admitted by the static isotropic plasma
equilibrium system \eqref{eq:PEE}, in spite of the equivalence (Theorem \ref{th:CGL_MHD}) between the isotropic and
anisotropic systems. Indeed, from \eqref{eq:finite_form_inf} it directly follows that for anisotropic equilibria, quantities
$\sqrt{1-\tau}\vec{B}$ and $p_\perp + \frac{1}{2}\tau {\bf{B}^2}$ are invariant with respect to $Y_{\infty}$. Hence,
according to the relations \eqref{eq:APEE2_1}, \eqref{eq:APEE2_2}, these infinite symmetries correspond to the
\emph{identity transformation} for isotropic plasma equilibria.

The structure and properties of infinite transformations \eqref{eq:finite_form_inf} and resulting families of solutions are
discussed below.

\subsection{Properties of the infinite symmetries \eqref{eq:finite_form_inf} of the static CGL
system}\label{subs_inf_prop}

The infinite symmetries \eqref{eq:sym_CGL_inf}, \eqref{eq:finite_form_inf} of the static CGL system  constitute a subgroup
of the infinite-dimensional symmetry group $G$ of dynamic CGL equilibrium system found in \cite{afc_ob}, and  share many of
their properties.

\noindent \textbf{(i) Structure of the arbitrary function.} The transformations \eqref{eq:finite_form_inf} depend on the
topology of the solution they are applied to, namely, on the set of magnetic field lines $\Psi=\const$. For the following
topologies of the initial solution, the domain of the function $\Psi=\Psi(\vec{x})$ is evident:

\begin{enumerate}
    \item The lines of magnetic field ${\bf{B}}$ are closed loops or go to infinity. Then the function $\Psi(\bf{x})$ is constant on each magnetic field line.
    \item The magnetic field lines are dense on 2D magnetic surfaces spanning the plasma domain $\mathcal{D}$. Then the function
$\Psi(\bf{x})$ has a constant value in on each magnetic surface, and is generally a function defined on a \emph{cellular
complex} (a combination of 1-dimensional and 2-dimensional sets) determined by the topology of the initial solution
$\{{\bf{B}},p_{\perp},p_{\parallel}\}$.
    \item The magnetic field lines are dense in some 3D domain $\mathcal{D}$. Then the function $\Psi(\bf{x})$ is constant in $\mathcal{D}$.
\end{enumerate}

\noindent \textbf{(ii) Topology, boundary conditions, and physical properties.} From \eqref{eq:finite_form_inf} it is
evident that ${\bf B}'\parallel{\bf{B}}$, thus  magnetic field lines of the original plasma configuration $\{{\bf
B},p_{\perp},p_{\parallel}\}$ are retained by the transformed solutions $\{{\bf B}',p'_{\perp},p'_{\parallel}\}$. Therefore
usual plasma equilibrium boundary conditions of the type ${\bf{n}}\cdot{{\bf{B}}}|_{\partial\mathcal{D}} = 0$ (${\bf{n}}$ is
a normal to the boundary ${\partial\mathcal{D}}$ of the plasma domain) are preserved.

If the free function $M(\Psi)$ is separated from zero, the transformed solutions retain the boundedness of the original
solution; the same is true about the magnetic energy ${{\bf{B}}^2}/2.$ For models in infinite domains, the free function
must be chosen so that new solutions $\{{\bf B},p_{\perp},p_{\parallel}\}$ have proper asymptotic behaviour at
$|{\bf{x}}|\rightarrow \infty$.

\bigskip \noindent \textbf{(iii) Stability of new solutions.} No general stability criterion is available
for MHD or CGL equilibria. However, several explicit instability criteria are known. In particular, under the assumption of
double-adiabatic behaviour of plasma {\cite{CGL}} the condition for the fire-hose instability is {\cite{cgl_instab}}
\begin{equation}
p_{\parallel} - p_{\perp} > {\bf{B}}^2, \label{eq:instab1}
\end{equation}
or, equivalently, $\tau>1$. According to  \eqref{eq:finite_form_inf}, we have
\begin{equation}
1-\tau_1=(1-\tau)/M^{2}(\Psi), \nonumber
\end{equation}
hence the infinite transformations \eqref{eq:finite_form_inf} \emph{do not change fire-hose stability/instability of the
original plasma configuration}.

The mirror instability {\cite{cgl_instab}} occurs when
\begin{equation}
p_{\perp} \left(\frac{p_{\perp}}{6 p_{\parallel}} - 1\right)
> \frac{{\bf{B}}^2}{2}. \label{eq:instab2}
\end{equation}
It can be shown that for every initial configuration $\{{\bf{B}},p_{\perp},p_{\parallel}\}$, there exists a nonempty range
of values which $M(\Psi)$ may take so that the mirror instability does not occur. The proof is parallel to that in
{\cite{afc_ob}}.

\bigskip \noindent \textbf{(iv) Symmetry group structure.}
We consider the set $G_C$ of all transformations \eqref{eq:finite_form_inf} with smooth $M(\Psi)$. Each such transformation
is uniquely defined by a pair $\{\alpha, H(\Psi)\}$:
\begin{equation}
M\left(\Psi\right)=\alpha \exp\left(H\left(\Psi\right)\right), ~~\alpha = \pm1. \nonumber
\end{equation}

The composition of two transformations $(\alpha, H(\Psi))$ and $(\beta, K(\Psi))$ is equivalent to a commutative group
multiplication
\begin{equation}
(\alpha,H)\cdot(\beta, K) = (\alpha\beta, H+K),\nonumber
\end{equation}
and the inverse $(\alpha, H)^{-1} =$ $(\alpha, -H)$. The group unity is $e=(1, 0)$. Thus $G_C$ is an abelian group
\begin{equation}
G_C = A_{\Psi} \oplus Z_2\label{new_group_st}
\end{equation}
with two connected components; $A_{\Psi}$ is the additive belian group of smooth functions in $\mathbb{R}^3$ that are
constant on magnetic field lines of a given static CGL configuration.

\section{Construction of anisotropic plasma equilibria}\label{sec:constr}
\setcounter{equation}{0}

In this section, we show how the infinite group of transformations \eqref{eq:finite_form_inf} is used to construct families
of anisotropic (CGL) plasma equilibria from a single known exact static CGL or MHD solution.

\subsection{Construction of general 3D anisotropic plasma equilibria}

\begin{theorem}[Construction of anisotropic plasma equilibria]\label{th:CGL_constr}
For any given solution $\{{\bf{B}},p_{\perp},p_{\parallel}\}$ of the static anisotropic (CGL) plasma equilibrium equations
\eqref{eq:APEE}, \eqref{eq:tau_cond}, or any given solution $\{{\bf{B}},P\}$ of the static isotropic (MHD) plasma
equilibrium equations \eqref{eq:PEE}, there exists an infinite family of anisotropic (CGL) plasma equilibrium solutions
given by \eqref{eq:finite_form_inf} depending on an arbitrary function of one variable. The infinite family of solutions has
the same set of magnetic field lines as the original solution.
\end{theorem}
This theorem follows directly from Theorem \ref{th:CGL_MHD} and infinite transformations \eqref{eq:finite_form_inf}.

In particular, for any given solution $\{{\bf{B}},P\}$ of the static isotropic (MHD) plasma equilibrium equations
\eqref{eq:PEE}, from \eqref{eq:finite_form_inf} we see that the corresponding infinite family of anisotropic (CGL) plasma
equilibrium solutions $\{{\bf{B}}',p_{\perp}',p_{\parallel}'\}$ is given by
\begin{gather}
{\bf{B}}'  = M(\Psi)\;{\bf{B}},~~\tau'=1-M^{-2}(\Psi), \nonumber\\
p_{\perp}'= P_1+P + \frac{1}{2}{\bf{B}}^2\left(1-M^{2}(\Psi)\right), ~~ p_{\parallel}'= P_1+P -
\frac{1}{2}{\bf{B}}^2\left(1-M^{2}(\Psi)\right), \label{eq:MHD_to_CGL_st}
\end{gather}
where $P_1$ is an arbitrary constant, $M(\Psi)$ is an arbitrary smooth function, and $\Psi=\Psi(\vec{x})$ enumerates
magnetic surfaces (or, in general, magnetic field lines) of the original isotropic plasma configuration.

\subsection{Axially and helically symmetric anisotropic plasma equilibria}

\bigskip \noindent \textbf{(i) Grad-Shafranov equation.} Bragg and Hawthorne \cite{BraggH} in 1950,
and Grad and Rubin {\cite{GradRubin}} and Shafranov {\cite{Shafr}} in 1958, have shown that the static isotropic (MHD)
system \eqref{eq:PEE} with axial symmetry (independent of the polar angle $\phi$) is equivalent to one scalar equation,
called Bragg-Hawthorne or \textit{Grad-Shafranov} equation (GS):
\begin{equation}\label{eq:GS}
\Psi _{rr} - \frac{\Psi _r }{r} + \Psi _{zz} + I(\Psi ){I}'(\Psi ) = - r^2 {P}'(\Psi ),
\end{equation}
Here $(r,\phi,z)$ are cylindrical coordinates, $\Psi(r,z)$ the unknown flux function (the function enumerating magnetic
surfaces), and pressure $P(\Psi)$ and function $I(\Psi)$ (related to poloidal magnetic field) is an arbitrary function.
Primes denote derivatives.

The expression for the magnetic field $\bf B$ is
\begin{equation}\label{eq:GS_B}
\vec{B} = \frac{\Psi _z }{r}\vec{e}_r + \frac{I(\Psi )}{r}\vec{e}_\phi - \frac{\Psi _r}{r}\vec{e}_z.
\end{equation}

\bigskip \noindent \textbf{(ii) JFKO equation.} Another reduction of static isotropic plasma equilibrium equations (\ref{eq:PEE}) describes \emph{helically symmetric}
plasma equilibrium configurations, i.e. configurations invariant with respect to the helical transformations
\[
z \to z + \gamma h,\,\,\,\varphi \to \varphi + h,\,\,\,r \to r.
\]
Equations \eqref{eq:PEE} in helical symmetry also reduce to one equation (Johnson-Frieman-Kruskal-Oberman, or \emph{JFKO}
equation \cite{jfko}) for one unknown function $\Psi(r,u)$, where $u=z-\gamma \varphi$ is the helical coordinate, and
$\gamma=\const\neq 0$:
\begin{equation}
\frac{\Psi _{uu} }{r^2} + \frac{1}{r}\left[ {\frac{r}{r^2 + \gamma ^2}\Psi _r } \right]_r + \frac{I(\Psi){I}'(\Psi)}{r^2 +
\gamma ^2} + \frac{2\gamma I(\Psi )}{\left( {r^2 + \gamma ^2} \right)^2} = - \mu {P}'(\Psi ). \label{eq:JFKO}
\end{equation}
At $\gamma=0$, \eqref{eq:JFKO} reduces to the GS equation \eqref{eq:GS}. Again magnetic surfaces are given by
$\Psi(r,z)=\const$, and $P(\Psi)$ is plasma pressure. The helically symmetric magnetic field is given by
\begin{equation}
{\bf B} = \frac{\Psi_u}{r}{\bf{e}}_r + B_\phi{\bf {e}}_\phi + B_z{\bf {e}}_z , \quad B_\phi = \frac{r I(\Psi) + \gamma
\Psi_r}{r^2 + \gamma^2}, \quad B_z = \frac{\gamma I(\Psi) - r\Psi_r}{r^2 + \gamma^2}.
\end{equation}

\bigskip \noindent \textbf{(iii) Axially and helically symmetric anisotropic (CGL) plasma equilibria.}
Though an analogue of the GS equation exists for anisotropic (CGL) plasmas \cite{anis_gs}, this equation is so complicated
compared to the original GS equation, that using it for finding exact solutions is impractical.

A practical way of obtaining families of axially and helically symmetric anisotropic plasma configurations \eqref{eq:APEE}
is through the application of transformations \eqref{eq:MHD_to_CGL_st} to known exact solutions of conventional GS or
JFKO equations (many of such solutions are known in literature, see e.g. \cite{ob_prl, jets4, ball2}). For a given axially
or helically symmetric isotropic (MHD) plasma equilibrium $({\bf B},P)$ in a domain $\mathcal{D}$, transformations
\eqref{eq:MHD_to_CGL_st} yield a family of anisotropic (CGL) equilibria. This family of solutions involves an arbitrary
function $M(\Psi)$ defined, in general, on a celluar complex, as follows (see Section \ref{subs_inf_prop}, part (i)):
\begin{itemize}
    \item In parts of the plasma domain $\mathcal{D}$ where magnetic field lines are dense on closed 2D magnetic surfaces,
    $M(\Psi)$ is a function of one variable $\Psi$ enumerating magnetic surfaces;

    \item In parts of $\mathcal{D}$ where magnetic field lines of the given MHD equilibrium are closed loops or go to infinity,
    values of $M(\Psi)$ can be chosen independently on each magnetic field line, i.e. $M(\Psi)$ is a function of two transverse coordinates.
\end{itemize}

In particular, starting from an axially (helically) symmetric isotropic plasma equilibrium \eqref{eq:PEE}, one may obtain a
family of axially (helically) symmetric anisotropic plasma equilibria \eqref{eq:APEE}, where in \eqref{eq:MHD_to_CGL_st}
$\Psi$ is the flux function solving the GS (JFKO) equation. Moreover, if the given isotropic plasma
equilibrium has magneric field lines that are closed loops or go to infinity, one may also obtain a wider class of
anisotropic plasma equilibria \eqref{eq:APEE}, by choosing the value of the arbitrary functions $M(\Psi)$ in
\eqref{eq:MHD_to_CGL_st} separately on every field line. In this case, the resulting anisotropic equilibria may have no
geometrical symmetries (\emph{symmetry breaking}).

\subsection{Example of an exact solution: an anisotropic plasma vortex.}\label{subs_examp_vort}
In \cite{ball4}, Bobnev derived a localized vortex-like solution to the static isotropic MHD equilibrium system
(\ref{eq:PEE}) in 3D space. The solution is presented in spherical coordinates $(\rho, \theta, \phi)$ and is axially
symmetric, i.e. independent of the polar variable $\phi$. \footnote{Bobnev's solution was found by an \emph{ad hoc} method
without using the GS equation. Since the solution is axially symmetric, it corresponds to some flux function $\Psi (r,z)$
that satisfies the GS equation. However, the explicit form of such flux function is not known.}

The solution has the form
\begin{equation} \label{eq:bob_sol1}
{\bf B}={\bf e}_{\rho}V(\rho)\cos\theta+{\bf e}_{\theta}W(\rho)\sin\theta+{\bf
e}_{\phi}U(\rho)\sin\theta,~~~P=P_0-p(\rho)\sin^2\theta,
\end{equation}
where
\begin{gather}
U(\rho) = \lambda_n \rho V(\rho),~~p(\rho) = \gamma \rho^2 V(\rho),~~W(\rho) = -V(\rho)-\rho V'(\rho)/2, \nonumber\\[1.4ex]
V(\rho)=B_0 \frac{V_0(2\lambda_n\rho)-V_0(2\lambda_n R)}{1-V_0(2\lambda_n R)},\nonumber\\[1.4ex]
V_0(x) \equiv 3\left(\frac{\sin x}{x^3}-\frac{\cos x}{x^2}\right),\nonumber\\[1.4ex]
\gamma=B_0 \frac{V_0(2\lambda_n R)}{1-V_0(2\lambda_n R)}=\const,~~~~B_0, P_0 = \const. \label{eq:bob_sol2}
\end{gather}

Here $\lambda_n,~ n=1,2,...$ is any member of the countable set of solutions of the equation
\begin{equation}
(3-4 R^2 \lambda_n^2) \sin(2R \lambda_n)-6 R \lambda_n \cos(2R\lambda_n)=0,~~R=\const.
\end{equation}
(In particular, $R \lambda_1\approx 2.882, R \lambda_2\approx 4.548, R \lambda_3 \approx 6.161$.)

Bobnev's solution satisfies boundary conditions
\begin{equation}
{\bf B}|_{\rho=R}=0,~~P|_{\rho=R}=P_0,~~{\bf B}|_{\rho=0}=B_0 {\bf e}_z,
\end{equation}
i.e. the domain ${W}\in \mathbb{R}^3$ where the magnetic field is nonzero is a sphere of radius $R$. The pressure outside of
${W}$ is constant: $P=P_0$. The magnetic surfaces $\Psi=\const$ inside ${W}$ are families of nested tori of non-circular
section, separated by spherical separatrices on which the pressure is $P=P_0=\const$; the number and mutual position of the
families depends on the choice of $R, \lambda_n$.

We take $R=1$, $\lambda=\lambda_3\approx 6.161$, $B_0=100$, and $P_0=4500$, and find $\gamma\approx-72.831$. Magnetic
surfaces $P=\const$ and levels of constant magnetic energy density ${\bf B}^2/2=\const$ (with $P$ and ${\bf B}$ given by
\eqref{eq:bob_sol1}) are shown in Figure 1 A, B respectively. The spherical separatrix magnetic surfaces have approximate
radii $\rho_1\approx 0.376,~\rho_2\approx 0.597.$


\begin{figure}[htbp]
\centerline{
\includegraphics[width=3in,height=3.2in]{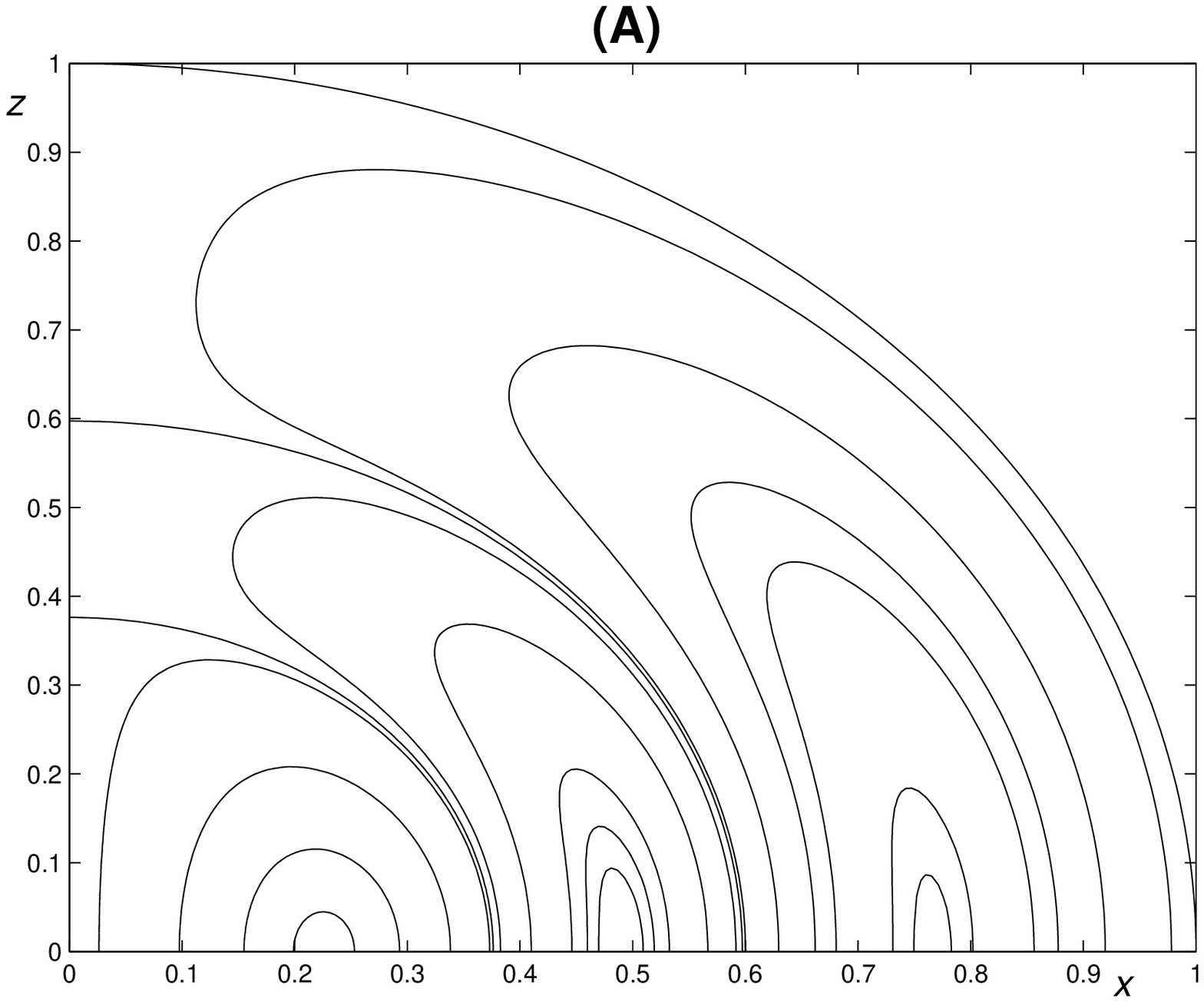}
\includegraphics[width=3in,height=3.2in]{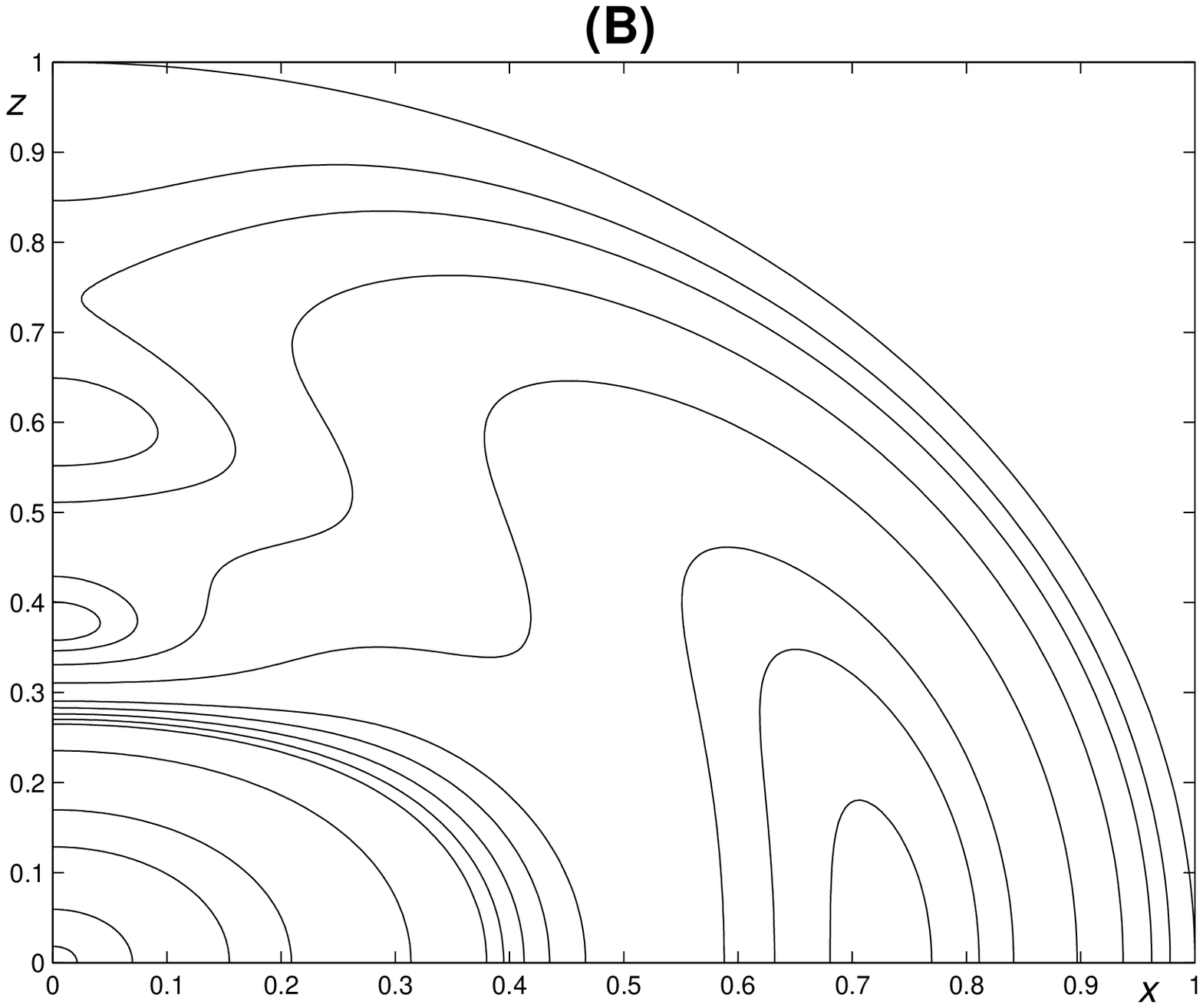}
}

\caption{ \label{fig1_1} Magnetic surfaces $P=\const$ (A) and levels of constant magnetic energy density ${\bf
B}^2/2=\const$ (B) in the static isotropic magnetic vortex \eqref{eq:bob_sol1} ($R=1, \lambda=\lambda_3 \approx 6.161$,
$B_0=100$, and $P_0=4500$). Here $z$ is the symmetry axis, and $x$ is the radial axis.}

\end{figure}

We apply transformations \eqref{eq:MHD_to_CGL_st} to this isotropic (MHD) solution, using the arbitrary function
\begin{equation}\label{eq:eg_M}
M(\Psi)=1+\frac{\Psi}{\Psi_1} \sin\left(\frac{\Psi}{\Psi_2}\right),~~~\Psi_1,\Psi_2=\const,
\end{equation}
which is constant on the magnetic surfaces of the original static MHD configuration given by \eqref{eq:bob_sol1}. The
arbitrary constants $\Psi_1,\Psi_2$ are chosen so that $M(\Psi)$ is separated from zero. [In the particular example below,
we take $\Psi_1=200, \Psi_2=60.$] As the result, we obtain an \emph{explicit anisotropic (CGL) plasma equilibrium
configuration} $\{{\bf{B}}',p_{\perp}',p_{\parallel}'\}$ given by \eqref{eq:MHD_to_CGL_st}, with the same set of magnetic
surfaces. The anisotropic pressure components $p_{\perp}'~,p_{\parallel}'$ are no longer constant on these surfaces. Contour
plots of $p_{\perp}'~,p_{\parallel}'$, and their profile along the radius of the vortex in the direction perpendicular to
$z$, are shown in Figure 2.


\begin{figure}[h]
\centerline{
\includegraphics[width=3in,height=3.2in]{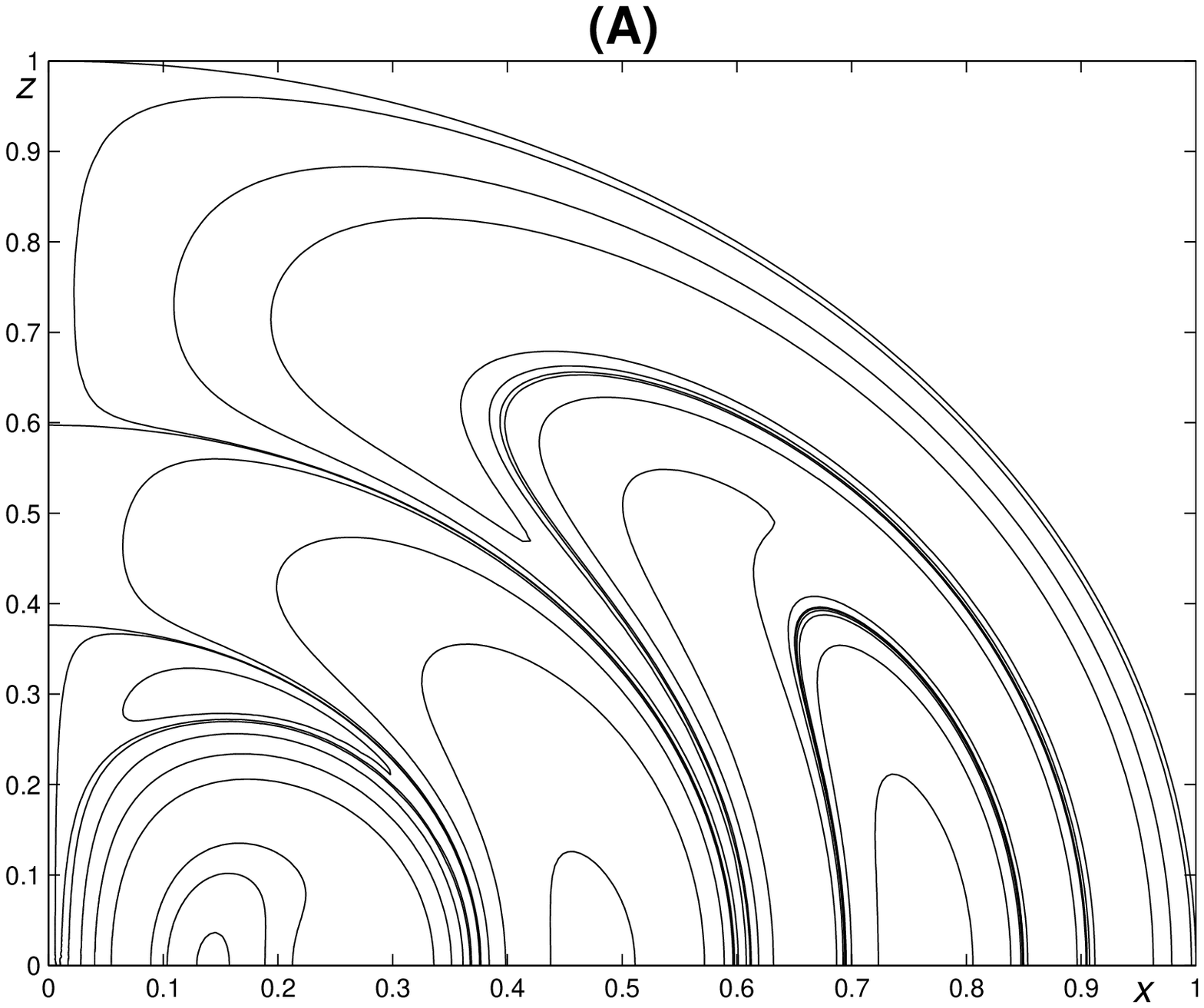}
\includegraphics[width=3in,height=3.2in]{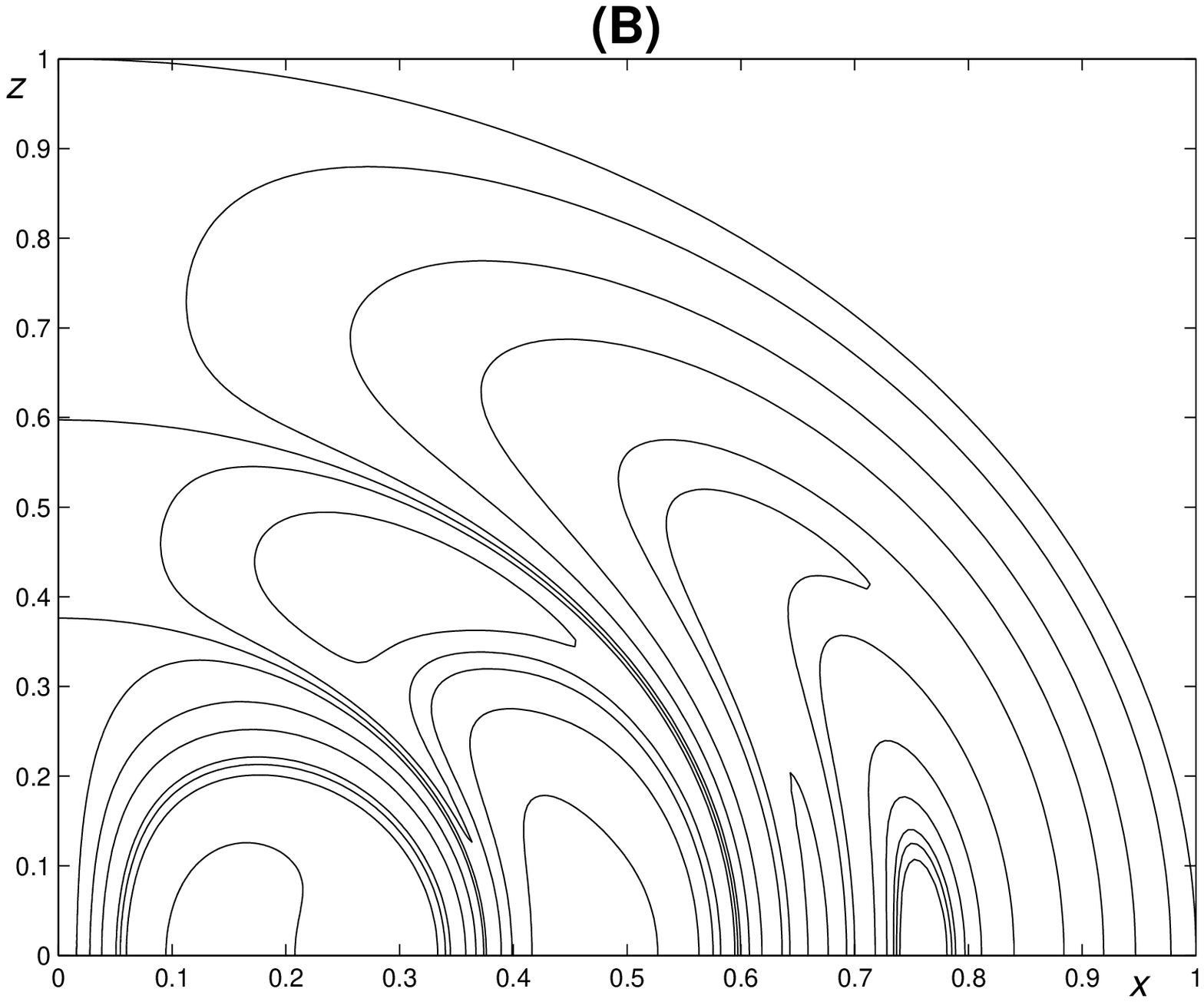}
}

\caption{\label{fig1_2} Surfaces of constant level of anisotropic plasma pressure components $p_{\parallel}$ (A) and
$p_{\perp}$ (B) in an anisotropic vortex ($R=1, \lambda=\lambda_3 \approx 6.161$, $B_0=100$, and $P_0=4500$). Here $z$ is
the symmetry axis, and $x$ is the radial axis.}

\end{figure}

Figure 3(A) shows profiles of magnetic energy densities $E={\bf{B}}^2/B_0^2$ and $E'={\bf{B}'}^2/B_0^2$ of the original
isotropic and the new anisotropic plasma vortex respectively. Figure 3 (B) shows profiles of normalized anisotropic pressure
components $p_{\perp}/B_0^2~,p_{\parallel}/B_0^2$ along the radius of the spherical vortex in the direction perpendicular to
the axis of symmetry $z$.


\begin{figure}[htbp]
\centerline{
\includegraphics[width=3in,height=3.2in]{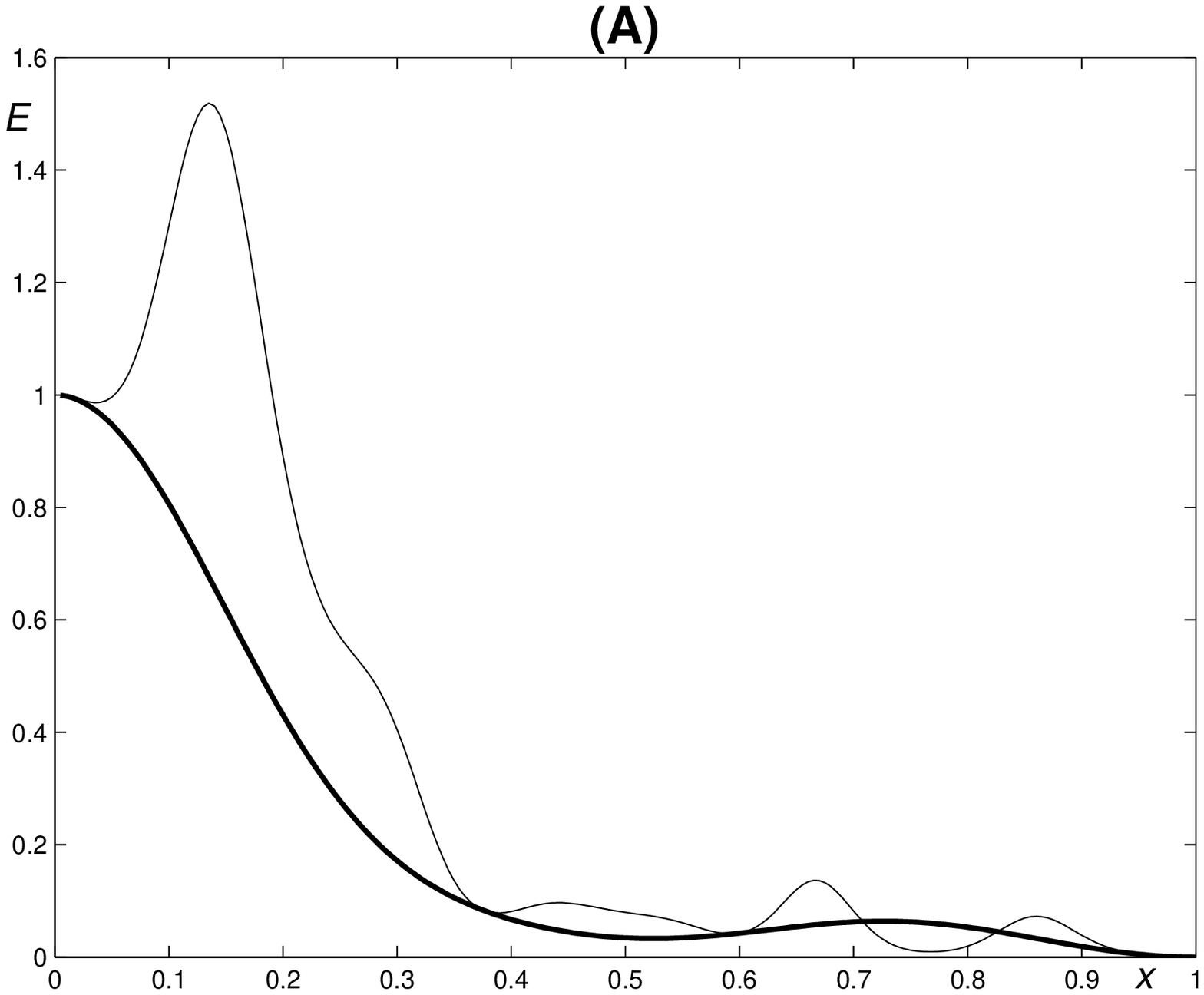}
\includegraphics[width=3in,height=3.2in]{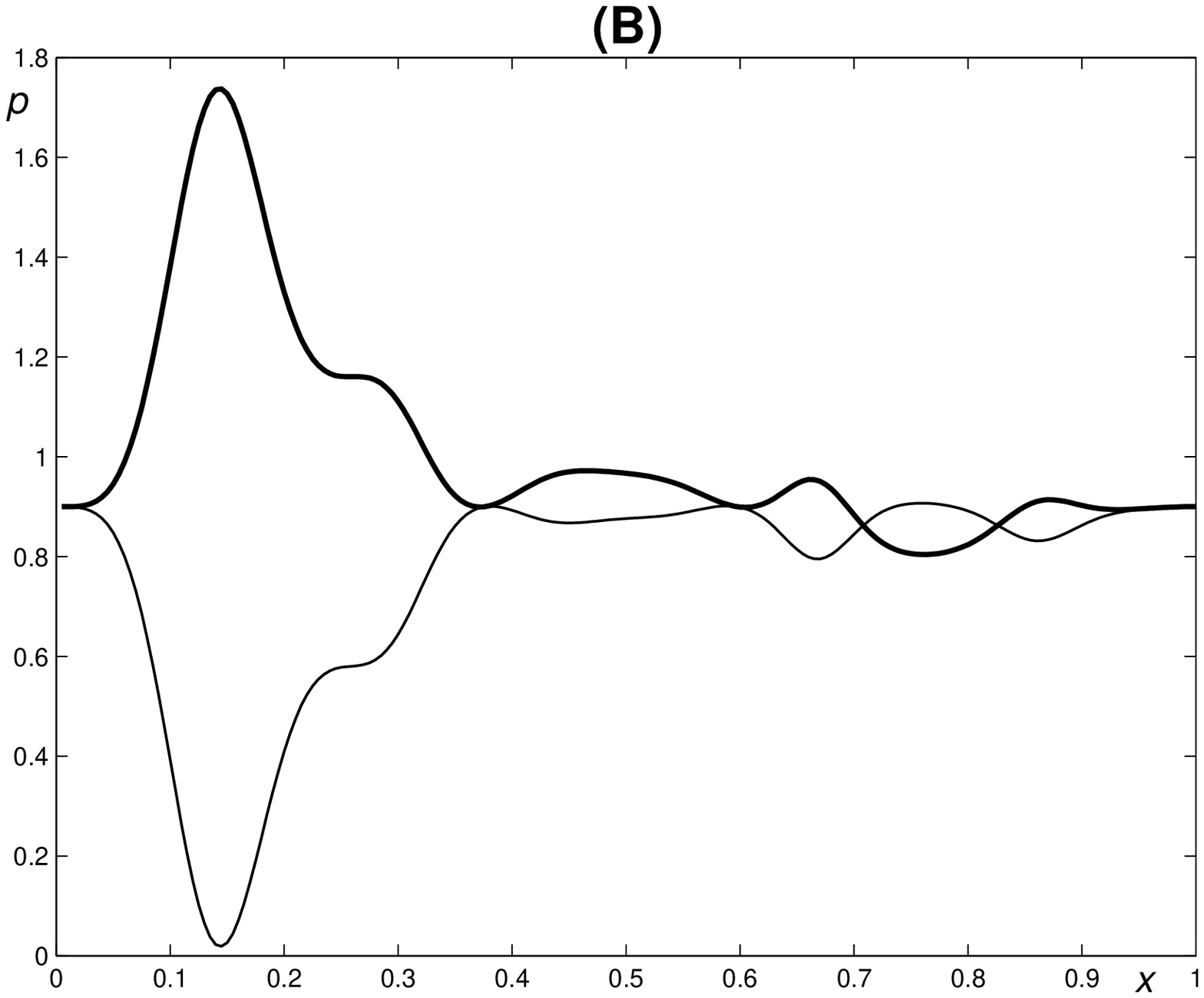}
}

\caption{\label{fig1_3} (A) Comparison of normalized energy densities $E={\bf{B}}^2/B_0^2$ (thick) and
$E'=({\bf{B}}')^2/B_0^2$ (thin) for isotropic and anisotropic plasma vortex. (B) Comparison of normalized plasma pressure
components $p_{\parallel}/B_0^2$ (thick) and $p_{\perp}/B_0^2$ (thin) in an anisotropic vortex in the radial direction. The
axis $x$ is chosen in the radial direction perpendicular to symmetry axis $z$. Here $R=1, \lambda=\lambda_3 \approx 6.161$,
$B_0=100$, and $P_0=4500$.}

\end{figure}

The presented solution is an explicit example of a physically meaningful axisymmetric static anisotropic plasma equilibrium
configuration in 3D space, arising from the model (\ref{eq:APEE}), (\ref{eq:tau_cond}). This solution is regular in the
whole domain (a ball of unit radius) and satisfies boundary conditions
\[
{\bf B}'|_{\rho=1}=0,~~p_{\perp}'|_{\rho=1}=p_{\parallel}'|_{\rho=1}=\const.
\]
The solution describes a static anisotropic plasma vortex confined by external gas pressure. Other choices of the arbitrary
function $M(\Psi)$ lead to different resulting anisotropic configurations.

\section{Conservation law analysis of static MHD and CGL equations}\label{sec:CL_analys}
\setcounter{equation}{0}

\subsection{Construction and interpretation of conservation laws}
A local conservation law is a continuity equation
\begin{equation}
\partial_t\Phi^0 +\div\vec{\Phi}=0 \label{eq:4d_cl}
\end{equation}
holding locally in 3D space for some physical variable. For dynamical plasma equations, one basic example is conservation of
mass where $\Phi^0=\rho$ is mass density and $\vec{\Phi}=\rho\vec{v}$ is momentum vector density in terms of the fluid
velocity $\vec{v}$. When static equilibria are considered, such conserved densities $\Phi^0$ are manifestly
time-independent, $\partial_t\Phi^0=0$, while the associated flux vectors $\vec{\Phi}$ are divergence free,
$\div\vec{\Phi}=0$. In general, a local conservation law of static MHD systems will be a vector density $\vec{\Phi}$ which
depends on the spatial coordinates $\vec{x}$, pressure $P$, magnetic field $\vec{B}$, and their partial derivatives with
respect to $\vec{x}$, such that it is divergence free for all static equilibria. Vector densities are physically trivial if
they identically have the form of curls, $\vec{\Phi} =\curl\vec{\Theta}$, when evaluated on static equilibria, with
$\vec{\Theta}$ being a local function of the same variables as $\vec{\Phi}$. A conservation law on static equilibria is thus
nontrivial if $\vec{\Phi}$ is not such a local curl, namely $\div\vec{\Phi}$ vanishes
essentially as a consequence of the static field equations. Conservation laws that
differ by a trivial one are considered to be physically equivalent.

The integral of a static conservation law in any domain $V$ in 3D space physically describes the net flux through the
boundary $\partial V$. In particular, if $V$ is a connected 3D region enclosed by a smooth closed surface $S$, then the net
flux
\begin{equation}\label{eq:net_flux}
\flux(\vec{\Phi},S) = \oint_S \vec{\Phi}\cdot d\vec{S}
\end{equation}
through $S$ vanishes on smooth static equilibria, while non-vanishing net flux for a static conservation law would indicate the presence
of a singularity in the flux $\Phi$ inside $V$. More generally, this net flux is independent of $S$, $\flux(\vec{\Phi},S_1) =
\flux(\vec{\Phi},S_2)$, due to Gauss' divergence theorem $\oint_{S_1} \vec{\Phi}\cdot d\vec{S} - \oint_{S_2} \vec{\Phi}\cdot
d\vec{S}= \int_V\div\vec{\Phi} dV=0$ where $V$ is the region bounded by the two surfaces $S_1,S_2$ in 3D space.

There is a computational algorithm (Direct Construction Method) for finding  conservation laws \eqref{eq:4d_cl}
\cite{olv,AB02p12,anco_scaling}. For static MHD systems, it is outlined as follows. Because static MHD systems are
Cauchy-Kovalevskaya type PDE systems (i.e. each system can be written in solved form with respect to a first partial
derivative of any one spatial coordinate), all of their nontrivial conservation laws arise from multipliers whose summed
product with the static field equations is identically a divergence. Specifically, for isotropic models without fluid flow
as considered hereafter, multipliers $\Gamma, {\bf \Lambda}=(\Lambda^1,\Lambda^2,\Lambda^3)$ are functions of the spatial
coordinates $\vec{x}$, pressure $P$, magnetic field $\vec{B}$, and their partial derivatives with respect to $\vec{x}$, such
that
\begin{equation}\label{eq:div_MHD}
\Gamma (\div\vec{B}) + \vec{\Lambda} \cdot (\curl\vec{B}\times\vec{B}-\grad P) = \div\vec{\Phi}
\end{equation}
holds identically (i.e. off of solutions).

Determining equations for multipliers can be obtained from the fact that divergences \eqref{eq:div_MHD} are annihilated by
variational derivatives (Euler operators) with respect to $\vec{B}$ and $P$. In the simplest case of multipliers with no
dependence on partial derivatives of $\vec{B}$ and $P$, the determining equations are equivalent simply to the adjoint of
the symmetry determining equations \eqref{eq:MHD_symm_deteq}:
\begin{equation}
\div\vec{\Lambda}=0,~~ \grad\Gamma = \vec{\Lambda}\times\vec{J}-\curl(\vec{\Lambda}\times\vec{B})
\end{equation}
holding for all static equilibria. For $\Gamma$ and $\vec{\Lambda}$ depending on partial derivatives of $\vec{B}$ and $P$,
there are additional determining equations involving  variational derivatives of the multipliers themselves. The
complete system of multiplier determining equations can be solved by an analog of Lie's algorithm for solving the symmetry
determining equations. When a set of multipliers is known, several approaches can be used to compute the flux vector
$\vec{\Phi}$ (for a comparison, see \cite{wolf_4met}). In particular, there exists an integral formula involving a homotopy
scaling of the field variables \cite{AB97, AB02p12}. Alternatively, since the static MHD system \eqref{eq:PEE} admits a scaling
symmetry \eqref{eq:sym_ScDil}a, the integral formula for $\vec{\Phi}$ can be replaced by a purely algebraic expression
\cite{anco_scaling} in terms of $\Gamma$ and $\vec{\Lambda}$.

\subsection{Conservation laws of the static isotropic plasma equilibrium system} \label{subs:CL_MHD}
For static MHD equilibria \eqref{eq:PEE}, we will determine nontrivial conservation laws. In particular, we seek multipliers
multipliers $\Gamma$ and ${\bf \Lambda}$, such that the summed product of these multipliers with the static MHD equilibrium
equations \eqref{eq:PEE} yields a nontrivial divergence
\begin{equation}\label{eq:MHD_mul_c_law}
\Gamma \div {\bf B}  +  {\bf \Lambda} \cdot \left(\curl~{\bf{B}}\times{\bf{B}} -\grad P\right) =\div {\vec{\Phi}}
\end{equation}
which vanishes when evaluated on static equilibria. An application of the Direct Construction Method \cite{AB02p12} leads to
the following results.

\begin{theorem}\label{th:CL_MHD}
The complete set of conservation laws admitted by the static isotropic plasma equilibrium system \eqref{eq:PEE}, for
multipliers $\Gamma,{\bf \Lambda}$ linear in first partial derivatives of $({\bf B},P)$ and with otherwise arbitrary
dependence on $(\vec{x},\vec{B},P)$, is given in Table \ref{table:CL_MHD}.
\end{theorem}

\begin{center}\renewcommand{\arraystretch}{1.6}\refstepcounter{table}\label{table:CL_MHD}
Table~\thetable: Conservation laws \eqref{eq:MHD_mul_c_law} of the static isotropic (MHD) plasma equilibrium system \eqref{eq:PEE} \\[2ex]\footnotesize
\begin{tabular}{|l|l|l|l|}
\hline \vspacebefore
\hfill \# $\hfill$ & \hfill Multipliers $\hfill$ & \hfill Conservation law $\hfill$& \hfill Remarks$\hfill$\\[0.5ex]
\hline \vspacebefore

1    &$\Gamma={\bf B}\cdot \vec{\zeta}$,~~ ${\bf \Lambda}=\vec{\zeta}$ & $\div(\vec{\zeta}\cdot\tens{T})=0$&
\mbox{\parbox[t]{7cm}{Conservation of stress and angular
stress, depending on a general Euclidean Killing vector $\vec{\zeta} =\vec{a}+\vec{b}\times\vec{x}$, with $\vec{a},\vec{b}\in \mathbb{R}^3$ constant vectors.\\}}\\[0.5ex]

\hline \vspacebefore

2    &$\Gamma=f(P),~\vec{\Lambda}=-f'(P)\vec{B} $ &  $\div (f(P)\vec{B}) = 0$ & \mbox{\parbox[t]{7cm}{Conservation of
magnetic flux depending on an arbitrary function that is constant on magnetic surfaces, which reflects the fact that
$P=\const$ on magnetic surfaces. \\}}\\[2.5ex]

\hline \vspacebefore

3    &$\Gamma=0, ~~\vec{\Lambda}=f'(P)\vec{J}$ & $\div ( f(P)\vec{J})=0$  &\mbox{\parbox[t]{7cm}{Generalized Kirchhoff's
current law. Reflects the fact that plasma electric current density ${\bf J}$
is tangent to surfaces of constant pressure. \\}}\\[0.5ex]

\hline
\end{tabular}
\end{center}
In Table \ref{table:CL_MHD}, the symmetric conserved tensor $\tens{T}$ is the sum of electromagnetic and fluid stress
tensors:
\begin{equation}\label{eq:t_tens}
\tens{T} =- \vec{B}\otimes\vec{B} +(P+\frac{1}{2}\abs{B}^2)\tens{I}.
\end{equation}

The arbitrary function $f(P)$ in conservation laws \#2 and \#3  has, in general, the form $f=f(\Psi)$ given by an arbitrary
function that is constant on magnetic field lines (and on magnetic surfaces, when they exist). Conservation law \#3
involving plasma electric current is analogous to conservation of vorticity in time-independent
incompressible Euler equations of fluid motion \eqref{eq:Euler}.

\subsection{Conservation laws of the static anisotropic plasma equilibrium system}
For static anisotropic (CGL) equilibria \eqref{eq:APEE}, \eqref{eq:tau_cond}, we likewise determine multipliers $\Pi$,
$\vec{\Omega}$, and $\Upsilon$, such that their summed product with the corresponding equations \eqref{eq:APEE},
\eqref{eq:tau_cond} is a nontrivial divergence:
\begin{equation}\label{eq:CGL_mul_c_law}
\Pi \div {\bf B}  +  \vec{\Omega} \cdot \left(\left(1-\tau\right) \curl{\bf B }\times{\bf B } - \grad~p_\perp
-\frac{1}{2}\tau\grad|\vec{B}|^2\right) + \Upsilon ({\bf B}\cdot \grad\tau)=\div {\vec{\Phi}}.
\end{equation}
Such divergence expressions vanish on anisotropic static equilibria and thus yield conservation laws.

The following theorem is obtained by an application of the Direct Construction Method \cite{AB02p12}.

\begin{theorem}\label{th:CL_CGL}
The complete set of conservation laws \eqref{eq:CGL_mul_c_law} admitted by the static anisotropic plasma equilibrium system
\eqref{eq:APEE}, \eqref{eq:tau_cond}, for multipliers $\Pi, \vec{\Omega},\Upsilon$ linear in first partial derivatives of
$({\bf B}, p_\perp, \tau)$ and with otherwise arbitrary dependence on $(\vec{x},{\bf B}, p_\perp, \tau)$, is given in Table
\ref{table:CL_CGL}.
\end{theorem}

\begin{center}\renewcommand{\arraystretch}{1.6}\refstepcounter{table}\label{table:CL_CGL}
Table~\thetable: Conservation laws \eqref{eq:CGL_mul_c_law} of the static anisotropic (CGL) system \eqref{eq:APEE}, \eqref{eq:tau_cond} \\[2ex]\footnotesize
\begin{tabular}{|l|l|l|l|}
\hline \vspacebefore
\hfill \# & \hfill Multipliers $\hfill$& \hfill Conservation law $\hfill$& \hfill Remarks$\hfill$\\[0.5ex]
\hline \vspacebefore

1    & \mbox{\parbox[t]{2.8cm}{$\Pi={(1-\tau)}({\bf B}\cdot \vec{\zeta})$,~~ \\$\vec{\Omega}=\vec{\zeta}$,\\
$\Upsilon=-{\bf B}\cdot\vec{\zeta}$}} & $\div(\vec{\zeta}\cdot\tens{S})=0$,& \mbox{\parbox[t]{6cm}{Conservation of stress
and angular
stress, depending on a general Euclidean Killing vector $\vec{\zeta} =\vec{a}+\vec{b}\times\vec{x}$, with $\vec{a},\vec{b}\in \mathbb{R}^3$ constant vectors.\\}}\\[0.5ex]

\hline \vspacebefore

2    & \mbox{\parbox[t]{3cm}{$\Pi=f,~~\vec{\Omega}=-f_p \vec{B},$\\$\Upsilon=f_\tau+\frac{1}{2} f_p |\vec{B}|^2,$\\$f=f(p,
\tau)$}} & $\div(f(p,\tau)\vec{B}) = 0$ & \mbox{\parbox[t]{6cm}{Conservation of magnetic flux depending on an arbitrary function that is constant on magnetic surfaces, which reflects the fact $p,\tau=\const$ on magnetic surfaces.\\}}\\[2.5ex]

\hline \vspacebefore

3    & \mbox{\parbox[t]{4cm}{$\Pi=0,~~\vec{\Omega}=f'(p)\vec{A},$\\$\Upsilon=-\frac{1}{2}f'(p)\vec{A}\cdot \vec{B};$\\
 here $\vec{A}=\curl \sqrt{1-\tau}\vec{B}$}} &$\div(f(p) \vec{A})=0$  &\mbox{\parbox[t]{6cm}{Conservation of flux related to vorticity of the vector field
$\sqrt{1-\tau}\vec{B}$. \\}}\\[0.5ex]

\hline
\end{tabular}
\end{center}
In Table \ref{table:CL_CGL}, similarly to the isotropic case, the symmetric conserved tensor $\mathbb{S}$ is a sum of the
electromagnetic stress tensor and the anisotropic fluid stress tensor:
\begin{equation}\label{eq:s_tens}
\tens{S} = -(1-\tau)\vec{B}\otimes\vec{B} +(p+\frac{1}{2}(1-\tau)\abs{B}^2)\tens{I}.
\end{equation}
The arbitrary function $f$ in conservation laws \#2 and \#3 is, in general, equal to a
 constant on magnetic field lines (and on magnetic surfaces, when they exist).

The classification of multipliers and fluxes admitted by the CGL equilibrium system in Table \ref{table:CL_CGL} coincides
with that of the isotropic MHD system in Table \ref{table:CL_MHD}, with the vector field $\sqrt{1-\tau}\vec{B}$
corresponding to the isotropic equilibrium magnetic field $\vec{B}$ and with $p$ corresponding to the isotropic pressure
$P$. Note that this relation directly manifests the equivalence of the systems stated in Theorem \ref{th:CGL_MHD}.

\subsection{Example 1: conservation laws of an isotropic plasma vortex \eqref{eq:bob_sol1}} \label{subs:CL_eg_Bob}

As a simple example, we consider integral forms of conservation laws for smooth static isotropic plasma equilibria (Table
\ref{table:CL_MHD}). If $V$ is a connected 3D region with a smooth boundary $S$, then integrals of conservation laws \#1,
\#2 and \#3 in Table \ref{table:CL_MHD} are respectively
\begin{equation}\label{eq:net_flux_eg}
\oint_S \vec{\zeta}\cdot\tens{T} \cdot \vec{n}\; dS = 0,\quad \oint_S f(P) \vec{B} \cdot \vec{n}\; dS =0,\quad  \oint_S f(P)
\vec{J}\cdot \vec{n}\; dS = 0,
\end{equation}
where $\vec{n}$ is an outer normal to $S$.

We now write down flux expressions for the Bobnev's isotropic plasma vortex solution presented in Section
\ref{subs_examp_vort}, when $S$ is a sphere of radius $\rho<R=1$. In spherical and cartesian coordinates, the arbitrary
constant vectors $\vec{a}, \vec{b}$ in a Euclidean Killing vector $\vec{\zeta} = \vec{a}+\vec{b}\times\vec{x}$ have the form
\begin{gather*}
\vec{a}=a_\rho\vec{e}_\rho+a_\theta\vec{e}_\theta+a_\phi\vec{e}_\phi = a_x\vec{e}_x+a_y\vec{e}_y+a_z\vec{e}_z,
\end{gather*}
\begin{gather}
a_\rho=a_x \sin\theta \cos\phi+a_y \sin\theta\sin\phi+a_z\cos\theta,~~~\nonumber\\
a_\theta = a_x\cos\theta \cos\phi+ a_y \cos\theta\sin\phi\-a_z\sin\theta,~~~a_\phi =
-a_x\sin\phi+a_y\cos\phi\label{eq:a_sph_cart}
\end{gather}
with similar expressions holding for $\vec{b}$. [Note that $a_x,
a_y, a_z$ and cartesian components of $\vec{b}$ are constants, whereas spherical components of $\vec{a}$ and $\vec{b}$
depend on spherical angles.]

Using the solution \eqref{eq:bob_sol1}, we find flux expressions
\begin{gather}
\vec{\zeta}\cdot\tens{T} \cdot \vec{n} =
a_\rho\left[-P_0+p(\rho)\sin^2\theta+\frac{1}{2}\left(V^2(\rho)\cos^2\theta-(U^2(\rho)+W^2(\rho))\sin^2\theta\right)\right]\nonumber\\
\quad\quad\quad\quad+ V(\rho)\sin\theta\cos\theta\left[U(\rho)(a_\phi-rb_\theta)+W(\rho)(a_\theta+rb_\phi)\right]; \label{eq:fluxes1_sph_eg}\\
f(P) \vec{B} \cdot \vec{n}= f(P) V(\rho)\cos\theta; \label{eq:fluxes2_sph_eg}\\
f(P) \vec{J}\cdot \vec{n}= 2f(P) \frac{U(\rho)}{\rho}\cot\theta. \label{eq:fluxes3_sph_eg}
\end{gather}

In particular, on each of the two separatrix spheres with radii $\rho=\rho_1, \rho_2$ (see  Section \ref{subs_examp_vort}),
one has $V(\rho)=0$, therefore $U(\rho)=p(\rho)=0$, $W(\rho)=-\rho V'(\rho)/2$, and $P=P_0=\const.$ Hence on these
separatrix spheres, fluxes \eqref{eq:fluxes2_sph_eg}, \eqref{eq:fluxes3_sph_eg} of the conservation laws \#2 and \#3 in
Table \ref{table:CL_MHD} vanish identically, and the flux \eqref{eq:fluxes1_sph_eg} of the conservation law \#1 becomes
\[
\vec{\zeta}\cdot\tens{T} \cdot \vec{n} = a_\rho\left[-P_0 +\frac{1}{2}\rho V'(\rho)^2
\sin^2\theta\right],~~\rho=\rho_{1},\rho_{2}.
\]
Substituting \eqref{eq:a_sph_cart}, it is easy to verify that the integral  $ \int_0^{2\pi} \int_0^{\pi}
(\vec{\zeta}\cdot\tens{T} \cdot \vec{n})\; \rho^2 \sin\theta \;d\theta d\phi$ vanishes.

\subsection{Example 2: conservation laws for axially symmetric plasma equilibria} \label{subs:CL_eg_2}

As a second example, we consider conservation laws admitted by static isotropic (MHD) equilibrium system \eqref{eq:PEE}
(Table \ref{table:CL_MHD}), and explicitly compute fluxes of these conservation laws for axially symmetric (Grad-Shafranov)
plasma equilibria \eqref{eq:GS_B} satisfying \eqref{eq:GS}.

Since the position vector in cylindrical coordinates $(r,\phi,z)$ is $\vec{x}=r\vec{e}_r+z\vec{e}_z$, the Euclidean Killing
vector takes the form $\vec{\zeta} = \vec{a}+\vec{b}\times\vec{x} = (a_r+ z b_\phi)\vec{e}_r+(a_\phi+b_z r- b_r
z)\vec{e}_\phi+(a_z-b_\phi r )\vec{e}_z$. Here
\begin{gather*}
\vec{a}=a_r\vec{e}_r+a_\phi\vec{e}_\phi+a_z\vec{e}_z=a_x\vec{e}_x+a_y\vec{e}_y+a_z\vec{e}_z, \\
\vec{b}=b_r\vec{e}_r+b_\phi\vec{e}_\phi+b_z\vec{e}_z=b_x\vec{e}_x+b_y\vec{e}_y+b_z\vec{e}_z,
\end{gather*}
are arbitrary constant vectors in $\mathbb{R}^3$;
\begin{gather*}
a_r=a_x\cos\phi+a_y\sin\phi,~~~a_\phi=a_y\cos\phi-a_x\sin\phi,\\
b_r=b_x\cos\phi+b_y\sin\phi,~~~b_\phi=b_y\cos\phi-b_x\sin\phi;
\end{gather*}
where $a_{x,y,z}, b_{x,y,z}=\const.$

Denoting the total plasma energy density $e=P(\Psi)+\frac{1}{2}\abs{B}^2$, and magnetic field components
\[
\vec{B} = B_r\vec{e}_r + B_\phi\vec{e}_\phi +B_z\vec{e}_z = \frac{\Psi _z }{r}\vec{e}_r + \frac{I(\Psi )}{r}\vec{e}_\phi -
\frac{\Psi _r}{r}\vec{e}_z,
\]
we find the following six conservation laws originating from conservation of stress and angular stress in Table
\ref{table:CL_MHD}:
\begin{gather}
\frac{1}{r}\frac{\partial}{\partial r}\left[r(e-B_r^2)\cos\phi + rB_r B_\phi \sin\phi\right]
-\frac{1}{r}\frac{\partial}{\partial\phi}\left[(e-B_\phi ^2)\sin\phi+B_rB_\phi \cos\phi\right]\nonumber\\
~~~~~~~ - \frac{\partial}{\partial z}\left[B_z(B_r\cos\phi - B_\phi \sin\phi)\right]=0,\label{eq:cl_nGS_1}\\
\frac{1}{r}\frac{\partial}{\partial r}\left[r(e-B_r^2)\sin\phi-rB_r B_\phi \cos\phi\right]
+\frac{1}{r}\frac{\partial}{\partial\phi}\left[(e-B_\phi ^2)\cos\phi-B_rB_\phi \sin\phi\right] \nonumber\\
~~~~~~~ - \frac{\partial}{\partial z}\left[B_z(B_r\sin\phi +B_\phi \cos\phi)\right]=0,\label{eq:cl_nGS_2}\\
\frac{1}{r}\frac{\partial}{\partial r}\left[rB_r B_z\right] - \frac{\partial}{\partial
z}\left[e-B_z^2\right]=0,\label{eq:cl_GS_1}
\end{gather}
\begin{gather}
\frac{1}{r}\frac{\partial}{\partial r}\left[r(z(e-B_r^2)+rB_rB_z)\sin\phi-rzB_r B_\phi \cos\phi\right]
+\frac{1}{r}\frac{\partial}{\partial \phi}\left[z(e-B_\phi ^2)\cos\phi+(rB_z-zB_r)B_\phi\sin\phi\right]\nonumber\\
~~~~~~~ - \frac{\partial}{\partial z}\left[(r(e-B_z^2)+zB_rB_z)\sin\phi +zB_\phi B_z\cos\phi)\right]=0,\label{eq:cl_nGS_3}\\
\frac{1}{r}\frac{\partial}{\partial r}\left[r(z(e-B_r^2)+rB_rB_z)\cos\phi+rzB_r B_\phi \sin\phi\right]
-\frac{1}{r}\frac{\partial}{\partial\phi}\left[z(e-B_\phi ^2)\sin\phi-(rB_z-zB_r)B_\phi\cos\phi\right]\nonumber\\
~~~~~~~ - \frac{\partial}{\partial z}\left[(r(e-B_z^2)+zB_rB_z)\cos\phi -zB_\phi B_z\sin\phi)\right]=0,\label{eq:cl_nGS_4}\\
\frac{1}{r}\frac{\partial}{\partial r}\left[r^2B_r B_\phi \right] + \frac{\partial}{\partial z}\left[rB_\phi
B_z\right]=0.\label{eq:cl_GS_2}
\end{gather}

For axially symmetric plasma configurations, the conservation laws corresponding to conservation of magnetic flux and
vorticity in Table \ref{table:CL_MHD} become trivial.

\begin{remark}
Conservation laws \eqref{eq:cl_GS_1} and \eqref{eq:cl_GS_2} have the axially invariant form $\frac{\partial}{\partial
r}\Phi_1(r,z) + \frac{\partial}{\partial z}\Phi_2(r,z)=0$ and are admitted by the Grad-Shafranov equation \eqref{eq:GS} for
\emph{any} choice of arbitrary functions $I(\Psi), P(\Psi)$. The other four conservation laws \eqref{eq:cl_nGS_1},
\eqref{eq:cl_nGS_2}, \eqref{eq:cl_nGS_3}, \eqref{eq:cl_nGS_4} are admitted by the full static MHD equilibrium system
\eqref{eq:PEE} but not by the Grad-Shafranov equation \eqref{eq:GS}, since they explicitly contain the angular variable
$\phi$.
\end{remark}

\section{Conclusions}\label{sec:further}

Magnetohydrodynamics (MHD) and Chew-Goldberger-Low (CGL) models are the two most widely used continuum plasma descriptions,
valid for the cases of isotropic and strongly magnetized (anisotropic) plasmas respectively. Knowledge of analytical
properties of these nonlinear PDE systems and, in particular, methods of finding exact solutions, are highly important for
applications.

In this paper we have classified complete sets of admitted point symmetries and conservation laws of the static isotropic plasma
equilibrium system \eqref{eq:PEE} and the static anisotropic plasma equilibrium system \eqref{eq:APEE}, \eqref{eq:tau_cond}.
This classification has led to establishing a direct transformation \eqref{eq:APEE2_1}, \eqref{eq:APEE2_2} between the two
systems. The transformation implies the equivalence of solution sets: every static anisotropic (CGL) equilibrium can be
obtained from a solution of static isotropic (MHD) equilibrium system, and to each solution of the static MHD system there
corresponds an infinite family of CGL equilibria depending on an arbitrary function defined on the set of magnetic surfaces.

The established equivalence yields an effective procedure of construction of exact explicit anisotropic (CGL) static plasma
configurations from a single known MHD or CGL solution. Many physically meaningful static MHD solutions are known, and each
of them gives rise to families of anisotropic (CGL) equilibria with the same topology of the magnetic field. It follows that
all axially and helically symmetric CGL configurations can be found from solutions of conventional Grad-Shafranov and JFKO
equations. An example of an explicit solution describing an anisotropic axially symmetric plasma vortex is given in Section
\ref{sec:constr}.

We note that symmetry classification of the static isotropic MHD system \eqref{eq:PEE} \emph{does not} lead to a symmetry
classification of the Grad-Shafranov or JFKO equations, since the latter are reductions of the system \eqref{eq:PEE}. The
conservation laws found in this paper for the static isotropic MHD system \eqref{eq:PEE} yield some particular conservation
laws of the Grad-Shafranov equation (Section \ref{subs:CL_eg_2}), but again not a full classification.

A natural next step will be to accomplish a similar complete symmetry and conservation law analysis of dynamic (${\bf v}
\neq 0$) plasma equilibrium models, and further, of time-dependent (non-equilibrium) equations. Dynamic equilibrium models
are already known to possess rich symmetry structure \cite{obsymm, afc_ob}, but the complete analysis was not done due to
the complexity of these PDE systems. Recently developed symbolic computation software \cite{GeM,wolf_soft} used in this
paper will be applied to study symmetries and conservation laws of these systems.

\bigskip \noindent \textbf{Acknowledgements}

A.F.C. is grateful to the Pacific Institute of Mathematical Sciences for postdoctoral research support; S.C.A. is supported
by an NSERC grant.



\end{document}